\documentclass[conference]{IEEEtran}
\usepackage{cite}

\ifCLASSINFOpdf
\else
\fi
\ifCLASSOPTIONcompsoc
 \usepackage[caption=false,font=normalsize,labelfont=sf,textfont=sf]{subfig}
\else
 \usepackage[caption=false,font=footnotesize]{subfig}
\fi

\usepackage[T1]{fontenc}
\usepackage{times}

\usepackage{booktabs}
\usepackage{multirow}
\usepackage{xspace}
\usepackage{enumerate}
\usepackage[shortlabels]{enumitem}
\usepackage{textcomp}
\usepackage{tikz}
\newcommand*\circled[1]{\tikz[baseline=(char.base)]{
    \node[shape=circle,draw,inner sep=0pt,minimum size=10pt] (char) {#1};}}

 \newcommand{\todo}[1]{}
 \DeclareRobustCommand{\added}[1]{#1}
 \DeclareRobustCommand{\removed}[1]{}

\newcommand{\sm}[0]{SMUFIN\xspace}
\newcommand{\mn}[0]{Marenostrum 3\xspace}

\begin{document}
%
\title{Accelerating K-mer Frequency Counting\\with GPU and Non-Volatile Memory}

\author{\IEEEauthorblockN{Nicola Cadenelli\IEEEauthorrefmark{1}\IEEEauthorrefmark{2},
Jord{\`a} Polo\IEEEauthorrefmark{1} and David Carrera\IEEEauthorrefmark{1}\IEEEauthorrefmark{2}}
\IEEEauthorblockA{
\IEEEauthorrefmark{1}Barcelona Supercomputing Center (BSC)\\
\IEEEauthorrefmark{2}Universitat Polit\`{e}cnica de Catalunya (UPC) - BarcelonaTECH
\\nicola.cadenelli@bsc.es, jorda.polo@bsc.es, david.carrera@bsc.es}
}
\maketitle

\begin{abstract}
The emergence of Next Generation Sequencing (NGS) platforms has increased
the throughput of genomic sequencing and in turn the amount of data that
needs to be processed, requiring highly efficient computation for its
analysis. In this context, modern architectures including accelerators and
non-volatile memory are essential to enable the mass exploitation of these
bioinformatics workloads. This paper presents a redesign of the main
component of a state-of-the-art reference-free method for variant calling,
\sm, which has been adapted to make the most of GPUs and NVM devices. \sm
relies on counting the frequency of \textit{k-mers} (substrings of length
$k$) in DNA sequences, which also constitutes a well-known problem for many
bioinformatics workloads, such as genome assembly.
We propose techniques to improve the efficiency of k-mer counting and to scale-up
workloads like \sm that used to require 16 nodes of \mn to a single machine with a
GPU and NVM drives.
Results show that although the single machine is not able to improve the time
to solution of 16 nodes, its CPU time is 7.5x shorter than the aggregate CPU
time of the 16 nodes, with a reduction in energy consumption of 5.5x.

\end{abstract}

\renewcommand\IEEEkeywordsname{Keywords}

\section{Introduction}

The field of computational genomics is quickly evolving in a continuous seek
for more accurate results but also looking for dramatic improvements in terms
of performance and cost-efficiency. Advances in next-generation sequencing
have been successful in lowering costs, enabling discovery of variants for
different kinds of diseases, and have also been widely adopted by the research
community.

These advances in genome sequencing technology make large-scale genomics
possible. However, mass exploitation of next-generation sequencing is still
computationally challenging since it requires dealing with ever growing
amounts of data and complex workloads. Moreover, many of these workloads have
been traditionally designed with a set of constraints that, on the
computational side limit flexibility and scalability, and on the biological
side limit accuracy.  For instance, many variant calling methods rely on
comparing sequences against a reference genome, which makes it harder to
accurately detect certain genomic variations.

Computing architectures are also evolving. The introduction of acceleration in
the form of GPUs and FPGAs, as well as Non-Volatile Memory, is enabling
methods that were unfeasible only years ago. To address the
aforementioned challenges and limitations the new emerging bioinformatics
workloads will need to leverage these modern computing architectures to remain
efficient and competitive.

One such example in the context of variant calling is \sm~\cite{smufin1}, a
state-of-the-art method that performs a direct comparison of normal and tumor
genomic samples from the same patient without the need of a reference genome,
leading to more comprehensive results. In its original implementation, this
novel approach required significant amounts of resources in a supercomputing
facility.

This paper presents how \sm has been adapted for efficient performance and
vertical scalability in a single node, exploiting GPUs and NVM (Non-Volatile Memory)
to accelerate one of its core components: counting the frequency of k-mers
(substrings of length $k$) in DNA sequences.
In particular, this paper describes the following main contributions:
(i) a way to construct large Bloom filters cooperatively between CPU and GPU;
(ii) a technique to minimize inter-thread communication by shuffling data
in the GPU; and
(iii) a customized mechanism to flush memory to non-volatile devices to
overcome memory capacity constraints.

The structure of the remaining sections of the paper is as follows.
Section~\ref{sec:background} introduces key computational genomics concepts, as
well as an overview of the foundations of \sm. Section~\ref{sec:method}
presents the acceleration method in detail.  Next, Section~\ref{sec:results}
shows results of the proposed changes.  And finally, Section~\ref{sec:related}
discussed related work, and Section~\ref{sec:conclusions} concludes.

\section{Background}
\label{sec:background}

A typical input of a genomics application consists of sequenced DNA samples
usually taking hundreds of GB.  Such samples are stored as heavily compressed
data and include short sequenced strings of DNA nucleobases called
\textit{reads}. Each sequenced genome sample typically contains $10^9$ to
$10^{10}$ reads, depending on some factors such as depth of coverage,
which indicates how many times each position in the genome is represented. The
length of each read is in the order of 10s to 100s of bases that are
represented by the four character alphabet $\{$\texttt{A}, \texttt{C},
\texttt{G}, \texttt{T}$\}$. Along with each base in a sequenced sample,
there's also an associated score that measures its quality.

\subsection{K-mer Frequency Counting}
\label{kmers}

Many genomics applications require splitting reads into smaller pieces called
\textit{k-mers}. Counting the frequencies of k-mers is widely used for genome
assembly and error detection, but it also has other applications such as
sequence alignment and variant calling.
k-mers of a nucleic acid read are all the possible sub-sequences within the
original read which have a length $k$.  The amount of k-mers in a read of
length $M$ is $M-k+1$. For instance, the number of 8-mers in a sequence of 10
bases is $10-8+1 = 3$, meaning \texttt{ACGGCAGCTG} has the following 8-mers:
\texttt{ACGGCAGC}, \texttt{CGGCAGCT}, and \texttt{GGCAGCTG}.
In addition to k-mers, \sm also defines the concept of \textit{stem}. A stem
is a fragment of a k-mer represented by its middle $k-2$ bases. That is,
removing the first and last bases from a k-mer. For instance, the stem
of the previous 8-mer is \texttt{-CGGCAG-}. Stems are used to group similar
k-mers and thus favor locality in the algorithm.

One of the challenges of k-mer counting when processing whole genome sequences
is data amplification. For a read of $M$ bases, and thus $M$ bytes, $M-k+1$
k-mers are generated.
For instance, a read of 100 bases requires at least
100 bytes, but for $k = 30$ its 71 k-mers take 568 bytes
(assuming 64 bits per k-mer -- 2 bits per each base).
This results in a 5.68x amplification factor that further increases when
lengthening the bases or shortening the k-mers.

\subsection{\sm}
\label{smufin}

The first implementation of the \sm method~\cite{smufin1} was based on
suffix trees. However, leveraging large suffix trees that inherently require a
locking mechanism to allow concurrent updates can be challenging. An
alternative implementation based on k-mers and hash tables to ease
parallelization and distribution of the workload is also available
and is the focus of the work presented in this paper.

The basic idea behind \sm can be summarized in the following steps:
(i) input two sets of nucleic acid reads, normal and tumoral;
(ii) build frequency counters of substrings in the input reads; and
(iii) compare branches to find imbalances, which are then extracted as
candidate positions for variation.

\sm is composed of ``checkpointable'' components called
\textit{units} that are combined to build fully-fledged workloads. Notice that
while units may resemble the individual programs that belong to a traditional
genomics pipeline, they are in fact part of a single application.  Units can
split data for processing into one or more \textit{partitions}, and each one
of these partitions can then be placed and distributed as needed: sequentially
in a single machine or concurrently in multiple nodes.
This kind of data
partitioning is achieved by reading the entire input multiple times,
discarding the parts that do not belong to the current partition.
In practice, scale-out executions with multiple lower-end nodes can run the
algorithm partitioning the input many times but duplicating IO.
Meanwhile,
at the opposite end of the spectrum, scale-up runs do not require as many
partitions, and hence less IO. Moreover, the application uses a second level
of distribution to further divide the work within each partition to multiple
threads; allowing each thread to work on a dedicated data structure without
requiring synchronization.

\begin{figure}
\centering
\includegraphics[width=1.00\columnwidth]{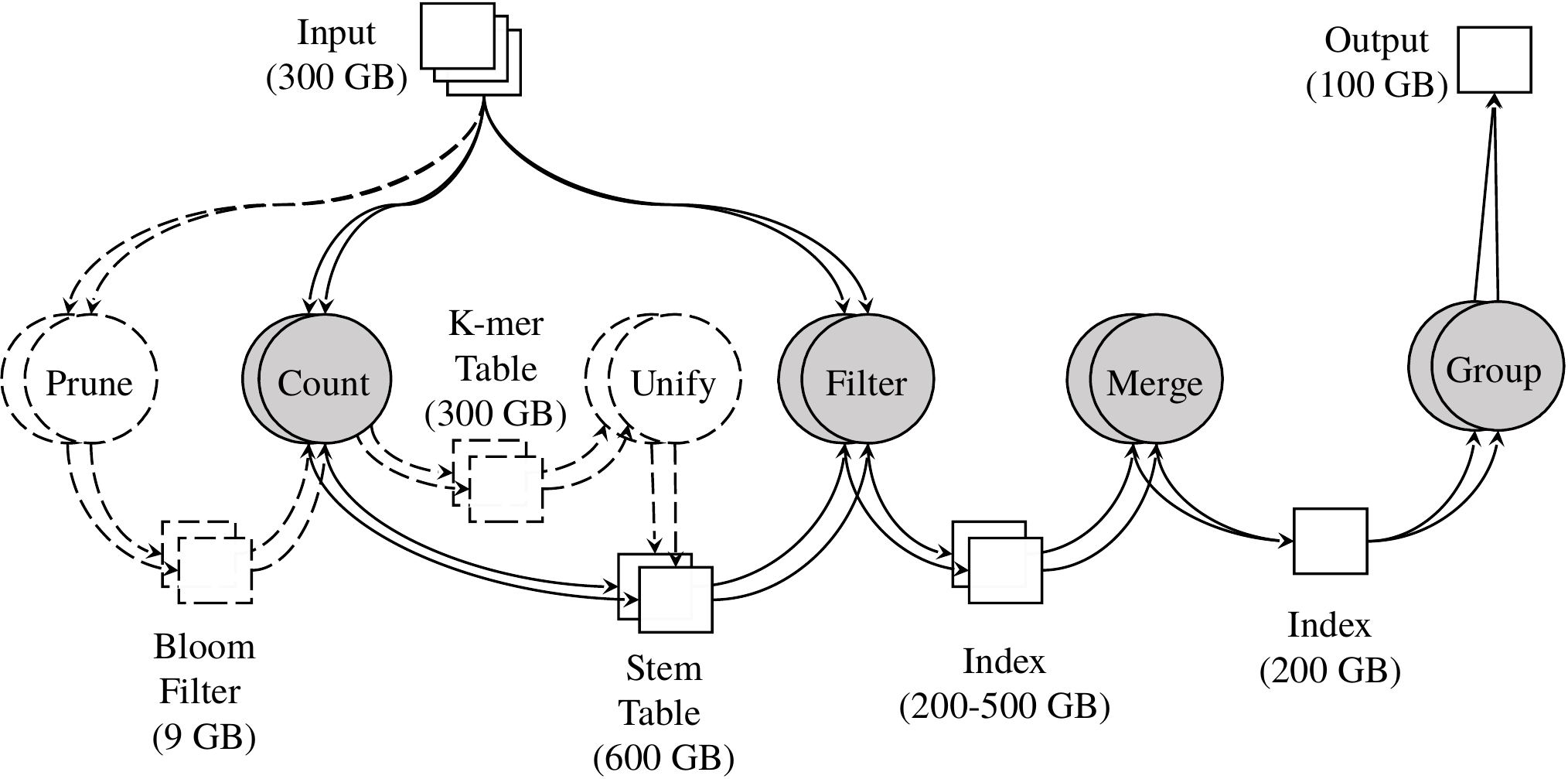}
\caption{\sm's variant calling architecture: overview of units and its data
flow (Section \ref{smufin}), along with proposed changes (Section
\ref{sec:method}).}
\label{f_overview}
\end{figure}

Figure~\ref{f_overview} provides an overview of the data flow between the main
units of \sm: \textit{Count}, \textit{Filter}, \textit{Merge} and
\textit{Group}; it also includes the proposed units introduced in this paper:
\textit{Prune} and \textit{Unify}, described in more detail in
Section~\ref{sec:method}. In this example, they are all configured to split
data into two partitions. As shown, three units read the entire
input one after another and generate intermediate results which are then
combined and assembled before producing the final output. The output of each
unit is read by the next one, either directly in memory, or serializing and
un-serializing to/from storage when there's not enough DRAM.

The goal and internal operations of each unit follow:
\begin{itemize}[leftmargin=*,itemindent=0em]
\item \textit{Count}: Builds a frequency table of normal and tumoral
    k-mers in the input sequences.  More specifically, k-mer counters
    are used to detect imbalances when comparing two samples, and it is
    designed to handle whole genome k-mers for values of $k$ in the range
    of $24 < k < 32$.
    While the number of k-mers in this range is potentially very large, \sm's
    variant calling algorithm only requires k-mers with a higher frequency in
    the tumoral sample.
    And because the natural k-mers distribution implies that most of them are
    seen only once, unique k-mers can be discarded.

\item \textit{Filter}: Selects k-mers with imbalanced frequencies, which are
    candidates for breakpoints or mutations. This selection is accomplished by
    reading the input sequences again and looking up its k-mers in the
    frequency table generated in the previous unit.

\item \textit{Merge}: Reads and combines multiple filter indexes from
    different partitions into single and unified indexes.

\item \textit{Group}: Matches candidate reads that belong to the same region.
    First, selecting reads that meet certain criteria, and second, retrieving
    other reads that share the same k-mers.

\end{itemize}

\section{Acceleration Method}
\label{sec:method}

The initial k-mer counting unit of \sm is one of the most computationally
demanding, and so it's the primary focus of the acceleration methods presented in
this paper. As illustrated in Figure~\ref{f_overview}, to improve the
performance of this unit, it has been changed and extended to include
two optional units: Prune and Unify. These units aim to reduce the main memory
requirement of the application when counting k-mers, which is the main
requisite when scaling up: the frequency table can take multiple TBs when
there is no data partitioning involved. Details of this extension follow:

\begin{itemize}[leftmargin=*,itemindent=0em]

\item \textit{Prune}: Adds all k-mers in the input files to a chain of two
    Bloom filters where only those already in the first (i.e., seen before),
    plus false positives, are propagated to the second one. At processing all
    the input, the second Bloom filter effectively constitutes a structure
    that can tell whether a k-mer has been observed more than
    once. This second filter is used in the Count unit to discard unique
    k-mers that are not relevant for the algorithm, reducing memory footprint
    and execution time.

\item \textit{Count}: As will be discussed in more details in~\S\ref{nvm},
    this new version changes the memory layout and how items are stored in the
    frequency table to keep a lower memory footprint. Changes also involve a
    custom mechanism to swap tables to an NVM drive, which also prevents running
    out of system memory when structures get too big.

\item \textit{Unify}: Combines swapped frequency tables in the new version of
    the Count unit and changes the memory layout to its expected form
    required by the following Filter unit. False positives given by the Bloom
    Filter are also removed at this point.
\end{itemize}

Note that the rest of the application was left unaltered and
the proposed changes are completely compatible with the original
implementation of the other units.

The following sections present a description of how the proposed changes have
been accelerated. First, describing how data encoding can be
offloaded, and presenting how to dimension data chunks for a double buffering
pipeline to keep the GPU as busy as possible while also overlapping data
transfer to and from the GPU. Next, we present how to shuffle data on the GPU
to minimize inter-thread communication between CPU threads, followed by a
discussion on how CPU and GPU can cooperate to build Bloom filters. Finally,
we illustrate how a manual swapping mechanism can be used to stage
data to an NVM device efficiently.

\subsection{Naive Offloading of CPU Intensive Operations}

To parallelize k-mer counting the work is split into two different
kinds of CPU threads: \textit{Loader} and \textit{Consumer}. The former threads
are committed to (i) load the compressed input files from storage to memory, (ii)
perform quality checks on the reads, (iii) generate and encode all the k-mers in
their 64-bit form, and (iv) send the k-mers to the Consumer threads. Consumer
threads, in turn, read incoming k-mers from the Loader threads and insert them
in the frequency tables. In the original version, communication between Loader
and Consumer threads happens via dedicated lock-free single producer single
consumer queues. This design has been improved to partially offload
computation to a GPU, including quality checking on the input DNA reads, and
the generation and encoding of all the k-mers to their 64-bit representation.

Efficient processing on the GPU requires splitting the $\approx$600 GB of
uncompressed input into smaller chunks. To overlap communication and computation of both
CPU and GPU a double buffering pipeline is used to stream data chunks to and
from the accelerator. This pipeline comprises of five stages that after a
ramp-up phase concurrently work on different data chunks at every cycle and where
the output of one stage is the input of the following stage in the
next cycle.
The stages can be summarized as:
(i) CPU Loader threads fill a host-side input chunk with DNA reads and quality
markers;
(ii) transfer an input chunk from host to accelerator memory;
(iii) the accelerator consumes an input chunk generating
all k-mers in an output chunk;
(iv) transfer an output chunk from accelerator to host memory; and
(v) CPU Consumer threads process the k-mers directly from the output chunk.
All synchronization is carried out by a new kind of CPU thread called
\textit{Orchestrator} thread.

The size of each chunk involved depends on the available GPUs and must be
carefully chosen to fully exploit the high degree of parallelism offered by
high-end GPUs. In particular, because each GPU thread (work-item) processes
one distinct DNA read; the number of reads in one input chunk has to be enough
to keep the processing elements of the accelerator busy.  Thus, the degree of
parallelism that the accelerator exhibits dictates the size of each chunk
requiring input chunks up to the order of hundreds of MB.  For what concerns
the output chunks instead, due to the amplification factor explained
in~\S\ref{kmers}, their size is simply a multiple of the input chunk.
Moreover, the usage of pinned memory allows a much better control on
data transfers and even simultaneous data transfers, one in each direction,
are possible when using GPUs equipped with two copy engines.

\subsection{Shuffling Data to Minimize Inter-Thread Communications}
\label{shuffling}

With a pipeline as described in the previous section, one of the challenges to
achieve efficient vertical scalability is to minimize data movement and communication.
Given a single output chunk coming from the GPU, all CPU Consumer
threads need to read the entire chunk {despite processing only parts of it.
Increasing data movement and constraining the scalability of the application.
On the other hand, also a queue-based approach similar to the one adopted
in the original version would limit scalability by adding hefty communication
between threads that exacerbates when increasing the number of Consumer threads.

To address this limitation, we designed an algorithm that shuffles
generated k-mers according to the CPU Consumer threads they belong to.
This approach also involves storing a list of offsets that allow Consumer
threads to know where exactly to start reading k-mers that belong to each
particular thread, effectively removing unnecessary communication to
distribute data to Consumers.

To shuffle the k-mers we implemented an algorithm, which resembles a parallel
Counting Sort that uses the data partitioning logic of \sm to select to
which Consumer thread each k-mer belongs but that does not sort the values within
buckets. Such algorithm was blended with the naive kernel that generates the k-mers.
The resulting algorithm is split into the following four kernels:

\begin{itemize}[leftmargin=*,itemindent=0em]
  \item{\textit{Zero-out kernel}: Resets all device-side data shared
  among different kernels from the previous cycle of the pipeline.}
  \item{\textit{Encode kernel}: Each GPU thread (work-item) generates all k-mers
  from a distinct DNA input read and count how many k-mers belong
  to each Consumer thread creating a histogram with as many bins as many Consumer
  threads.
  GPU threads cooperate (using atomic add operation) to build,
  first, a local histogram (at work-group scope) and then a global histogram.
  }
  \item{\textit{Prefix-sum kernel}: Performs an exclusive prefix-sum on the
  global histogram.
  Note that the result of the prefix-sum is the list of offsets that tells to each
  Consumer thread how many k-mers and where to start reading them in the output chunk.
  }
  \item{\textit{Broker kernel}:
  Per each Consumer thread, each work-group copies the, already locally shuffled,
  k-mers to the output buffer applying the offsets obtained from the previous
  prefix-sum; shuffling all k-mers by the Consumer threads.}
\end{itemize}

The main reason why we split the algorithm into four kernels is that it is the
only way to obtain a global synchronization point among GPU threads.
Note also that extra GPU-side buffers are required to share data between kernels
and to store k-mers in the Encode kernel.
Whereas the former is a common practice of GPU programming and occupies tens of MB,
the latter is a requirement of the algorithm and must be of the same size of an
output chunk.
Such algorithm is in fact not an in-place algorithm, and if we were to use the
same buffer, some threads might move some k-mers before other threads even start.
Thus, producing erroneous results.

Such solution considerably reduces CPU-side data movement enabling
vertical scalability to more CPU threads at the expense of extra computation and
more memory consumption on the GPU side.

\subsection{Cooperative CPU-GPU Construction of Very Large Bloom Filters}

As explained earlier in this Section the second main contribution of this work
is part of the Prune unit, which is used to build a chain of two Bloom filters
that allow discarding k-mers that are seen only once.

Whereas in the Count unit the offloading to the GPU of the lookups to the second
level Bloom filters is natural and allows to further unburden the Consumer threads.
In the Prune unit, we are populating the Filters on the CPU side so the same cannot be done.
However, when building the chain of Filters, if an item is already in the second
level adding it again is superfluous.
Hence, while building the chain of Bloom filters the second level can be
offloaded to the GPU to drop those k-mers already in the list.
This opposite behavior from the Count unit reduces the number of k-mers that
must be processed by the Consumer threads in the Prune unit.
Note also that, in the Prune unit the GPU copy of the second level Bloom filter
must constantly be updated at every cycle of the pipeline.
Which increases CPU to GPU communication.

\subsubsection{Overcoming GPU Memory Constrains}
\begin{figure*}
\includegraphics[width=1.00\textwidth]{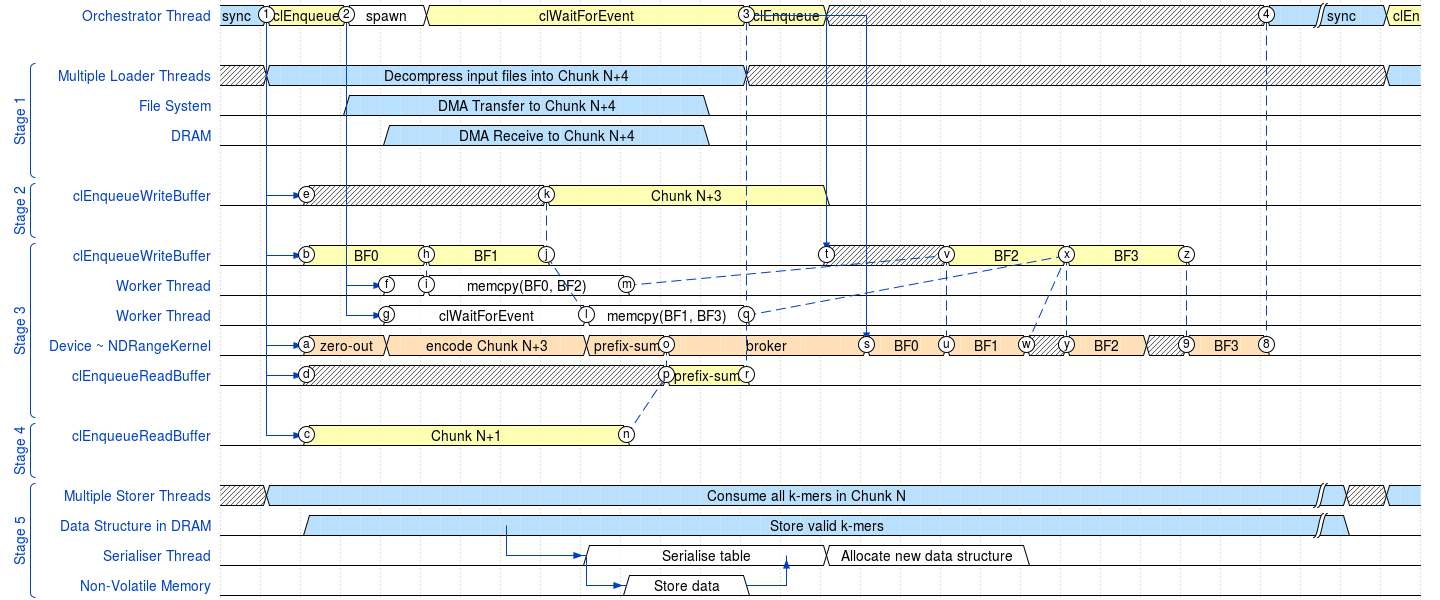}
\caption{Activity involved at every cycle of the pipeline.}
\label{f_pipeline2}
\end{figure*}

Albeit straightforward to implement, offloading the second level of Bloom filter
poses a challenge when aiming to remove data partitioning.
Such filter is in fact, of around 9 GB which due to the pinned buffer might
even require twice as much in the GPU memory.
Characteristic not common neither in high-end GPUs.

However, after the Broker kernel, when the k-mers are shuffled per Consumer
thread, the lookups to the Bloom filter can be split in multiple Bloom filter
kernels. One per each Consumer thread accessing a copy of the Bloom filter of
the relative CPU thread.
And to relax the requirement on the device memory, we added to our pipeline the ability to
virtualize the GPU memory in DRAM.
To do this, the Orchestrator thread allocates as much as
possible GPU pinned buffers of the size of each second Bloom filter.
Plus, in DRAM it also allocates enough memory to store the Bloom filters.
And, just before to execute a Bloom filter kernel, the Orchestrator thread
copies the Bloom filter to the GPU memory.
In Figure~\ref{f_pipeline2} we show the activity involved in one cycle of the pipeline
and the coordination work done by the Orchestrator thread can be summarized as follows:
\begin{enumerate}[(I)]
  \item Enqueue \circled{1} an asynchronous data transfer to read the global prefix-sum as
  soon as the prefix-sum kernel completes \circled{o}.
  \item Enqueue \circled{1} asynchronous data transfers \circled{b} and
  \circled{h} to write the first pinned
  buffers to the accelerator.
  \item Spawn \circled{1} CPU worker threads that wait \circled{f} for the
  pinned buffer transfer to complete \circled{h} and copy the next stage buffer
  to the pinned buffer \circled{i}.
  \item Wait \added{for }the transfer of the prefix-sum \circled{4} and enqueue the first round
  of Bloom filter Kernels \circled{s}.
  \item Enqueue asynchronous transfers of next Bloom filters \circled{t}.
  These transfers will start only after the relative worker thread finished
  the copy \circled{m} and the Bloom filter kernel completed \circled{u}.
  \item Restart from (III) until all Bloom filters are copied to the
  pinned buffer, copied to the accelerator and its relative kernel executed.
  \item As soon last Bloom filters are copied, and while executing last Bloom
  Filter kernels, spawn worker threads to overwrite the first pinned buffer
  for next cycle of the pipeline (not in Figure~\ref{f_pipeline2}).
\end{enumerate}

Note that our mechanism is similar to what OpenCL 2.0 Shared Virtual Memory
and CUDA Unified Memory offers. However, some GPU vendors do not support
OpenCL 2.0. Moreover, even if OpenCL 2.0 Shared Virtual Memory would be available
our mechanism can be simplified but it would still be valid to prevent GPU-side
page faults.

\subsection{Custom Swapping to NVM}
\label{nvm}

The main reason for the Prune unit is to reduce the amount of DRAM required in
the Count unit providing a small Bloom filter to drop all those k-mers seen only
once in input.
However, even with the Prune unit, the frequency table used in the Count unit
requires hundreds of GB.

To relax this requirement, we implemented a mechanism for which,
accordingly to the amount of system DRAM,
we set a maximum size for the table of each Consumer thread
and, once full, a table is swapped to the memory extension by a new thread while
an empty table, previously allocated as ``hot spare'', becomes active. To
reduce the overall data written the size of each table is recommended to be
as big as possible since the bigger it is, the more likely repeated k-mers
will be seen again.
Thus, leading to fewer duplicate elements in the tables swapped to the NVM device.
Moreover, writing big chunks to non-volatile memories allows exploiting internal
parallelism typical of flash drives~\cite{chen_essential_2011}.

This custom swapping mechanism substitutes the OS swapping, which is disabled
to prevent performance deterioration due to kernel-jittering. To
avoid the overhead of a file system software stack and to have a
byte-addressable memory, we access the NVM drive as a block device
which gets memory mapped and addressed as normal memory.
We also developed a bookkeeping system to store all metadata required to
identify and retrieve each swapped table.
Such data comprise, but is not limited to, information
about: the partitions to which a table  belongs, if a table is ``stale''
and can be removed or not, the offset in bytes from the beginning of the NVM drive,
and the number of items contained in each table.
Metadata is kept in main memory and is written to a reserved area
at the beginning of the memory extension right before
un-mapping the drive.

Moreover, now that we store the tables to the memory extension and we
merge them in the Unify unit we can adopt different data layouts from Count to
Filter unit to address the needs of both units.
In details, in the original implementation, the data layout was chosen to provide
maximum data locality in the Filter unit where given a k-mers all counts of all
k-mers belonging to the same stem are checked. For this reason, counts are
grouped per stem creating items of 72 Bytes -- 8 for the stem and 2 per each of
the 16 normal and 16 tumoral k-mers counts.
However, a better layout for the Count unit is to store counts per k-mer in 16
Bytes -- 8 for the k-mer, 4 for the two counts and 4 of padding added
by the \textit{std::pair\textless key,item\textgreater} used in the Google's
sparse hash table used in \sm.
A layout that, given the many zeros in the stem form, reduces the memory
footprint of the overall tables and also limits the memory wasted for each false
positive given by the Bloom filters to only 16 Bytes.

Google's hash maps implementation
allows the serialization of a table to a stream which we could have used to swap
and load tables back to memory.
However, while this constitutes a neat way it has two issues
that increase the amount of data required to store a serialized table.
The first is that this call also stores metadata to ensure that once a table is
loaded back to memory it will have each item at the same position
of the same bucket as in the original table which is something we do not need.
The second issue instead, is that it copies each
\textit{std::pair\textless key,item\textgreater}
as it is, including eventual padding used for data alignment within the
\textit{std::pair}.
If for small volumes these two issues might not be a problem.
In our tables of k-mers, this extra padding, which as explained above is 25\% of
the memory footprint, amounts to at least 100 GB.
Hence, it should be prevented.
As a solution, we implemented a similar routine which loops
throughout all items and copies both key and item from the
\textit{std::pair\textless key,item\textgreater},
skipping the padding, to a memory region which can be flushed thereafter.

To reduce IO bursts given by the serialization of multiple tables at the same time
the application uses slightly different maximum sizes from table to table.
Furthermore, the swapping mechanism can be tuned to use only a few working threads
to reduce the number of context switches when the CPU is already
saturated by the rest of the application.

Note that this proposed manual swapping mechanism also allows the use of
Google's dense hash table that, compared to the sparse implementation,
offers considerable better performance at the expense of a higher memory footprint.

\section{Results}
\label{sec:results}

In order to evaluate the impact of the presented work, we compare execution
time and energy consumption to run \sm in two different environments.
First, a large supercomputer with many distributed cores and
a high-speed interconnect where \sm is usually deployed in production.
Second, a single customized node with specialized
hardware has been used to evaluate the vertical scalability.
Note that even if the work involved only the initial unit of k-mer counting,
the benefit of reducing the number of partitions and using fast local
storage can also be seen in the following units.
Due to this, we present results relative to the entire \sm application.
The evaluation concludes with an analysis of bottlenecks and limitations of the
scale-up system used.

\subsection{Evaluation Methodology}
The distributed evaluation has been executed in 16 nodes of \mn,
a supercomputer based on
Intel SandyBridge-EP running SuSe Distribution 11 SP3. Each node used was
equipped with 2x 8-core E5-2670 2.6GHz, one 500GB 7200 rpm SATA II local disk,
and 8x 16G DDR3-1600 DIMMs for a total of
128 GB of main memory.
Each rack is composed of 84 compute nodes and
2 BNT RackSwitch G8052F that connects to a 1.9 PB of GPFS disk storage.
Vertical scalability is instead tested in one machine equipped with: two Intel Xeon
CPU E5-2680v3 @ 2.50GHz, one Nvidia Tesla K40c, sixteen 32-GB DDR4 DIMMs
running at 2133 MHz for a total of 512 GB of DRAM, one FusionIO SX350-3200
used as local storage, and one FusionIO
SX350-1600 used as memory extension.
The software stack is composed by Ubuntu
16.10 with a 4.4.0-72-generic kernel, Nvidia Driver version 375.39
offering OpenCL 1.2 version, and FusionIO driver 4.3. The GPU is set with
ECC enabled and GPUBoost disabled.
Moreover, to ascertain the effectiveness of the GPU we tested this
configuration with the GPU and without it.
The number of CPU threads changes accordingly to the number of cores
in each machine. For example, in each of the 16 nodes of \mn we used 8 loaders and
8 Consumers, whereas in
the scale-up machine we increased the number of Consumer threads to 48.
In both machines, we process the same personalized genome based on the Hg19 reference,
with randomly chosen germline and somatic variants as described
in~\cite{smufin1}, including SNPs, SNVs (more than 100 bp apart),
translocations, and random insertions, deletions and inversions, all ranging
from 1 to 100Mbp. In silico sequencing was simulated using ART Illumina21. The
total size of the final normal and tumoral samples is 312GB of gzip compressed
FASTQ files.
In \mn, each of the 16 nodes takes care of one different data partition.
On the scale-up version instead, thanks to the solutions proposed in this work, data
partitioning is not required for counting k-mers.
However, the frequency table in the stem format occupies around 600 GB;
requiring two partitions starting from the Filter unit.

\begin{figure}
\centering
\includegraphics[width=1.00\columnwidth]{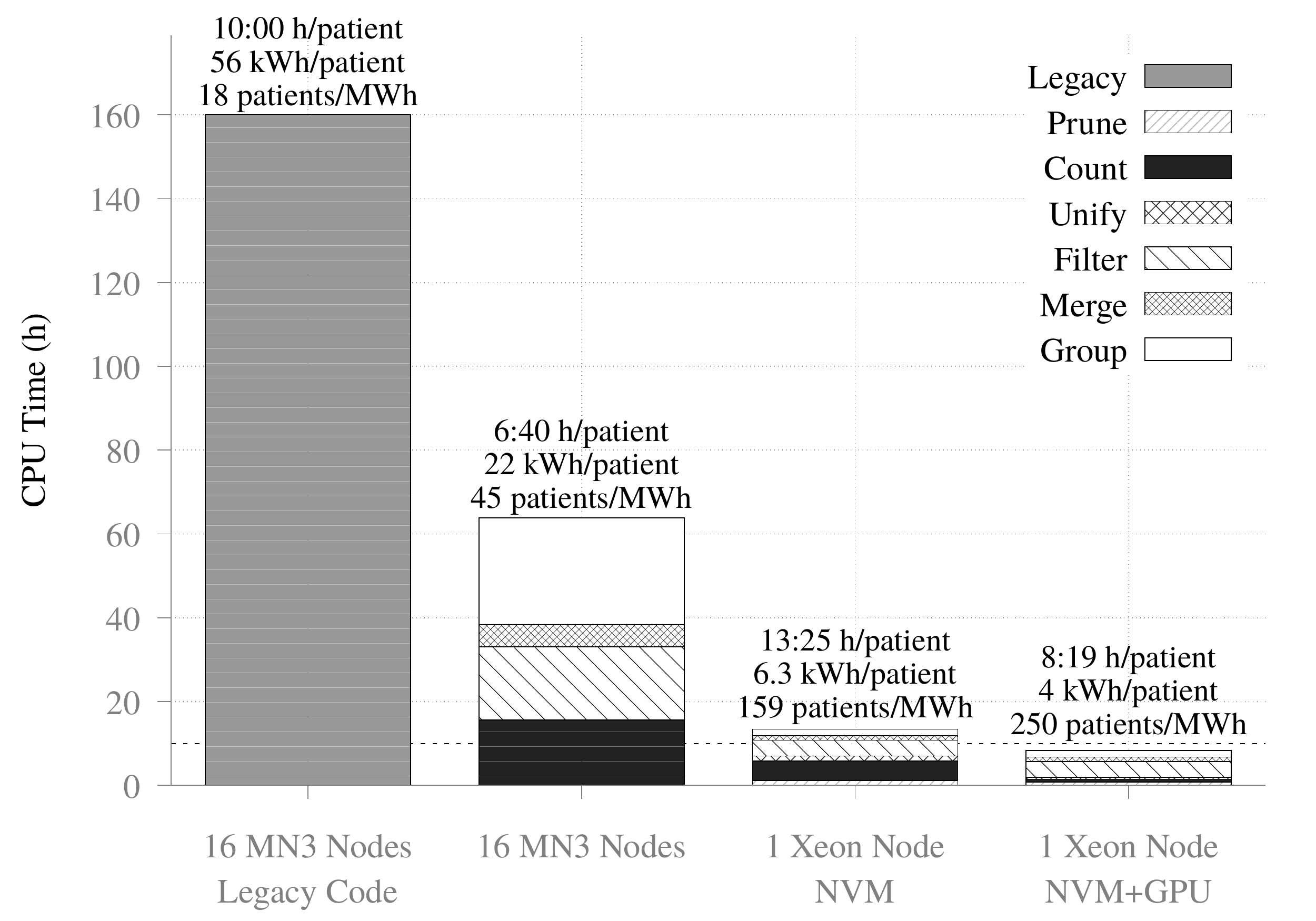}
\caption{Aggregate CPU time, time to solution and energy consumption of \sm running in
16 \mn nodes and in 1 Xeon-based node with NVM drives without and with the GPU.
Energy
consumption was collected using IPMI (Intelligent Platform Management Interface).
}
\label{f_time}
\end{figure}

\subsection{Performance and Energy Consumption}
Figure~\ref{f_time} shows the execution time and energy consumption of \sm on both environments.
Note that to ease comparison of the results all times related to the
\mn are aggregated as if the execution of all data partitions had been performed
sequentially. The dashed horizontal line in the chart marks the execution time
of the legacy code in parallel and shows that the scale-up machine using the GPU
outperforms the 16 nodes of \mn when running the legacy code based on suffix trees.
The figure also reveals that units not involved in this work -- Filter, Merge,
and Group -- have significant better performance on our single customized node
compared to the distributed environment. Such difference is only in part due to
the diverse CPU generations between systems and it's mainly due to the reduced
number of data partitions and to the use of local NVM which
helps IO bound units like Merge and Group.
The results show that even though both scale-up configurations have higher time to solution
than the 16 \mn nodes, they are clearly better in terms of CPU time and energy consumption.
In particular, the scale-up configuration with the GPU has a CPU time 7.5x shorter
than the distributed while also reducing the energy consumption of 5.5x.
The benefits of using a GPU are only relative to k-mer counting and
can be seen comparing the single node execution without and with the GPU.
And while with the GPU the overall time to solution is 1.67x shorter than without
the GPU, the improvement relative to only the k-mer counting is instead of 3.67x.

\subsection{Characterization of the Accelerated Version}

\begin{figure}
\includegraphics[width=1.00\columnwidth]{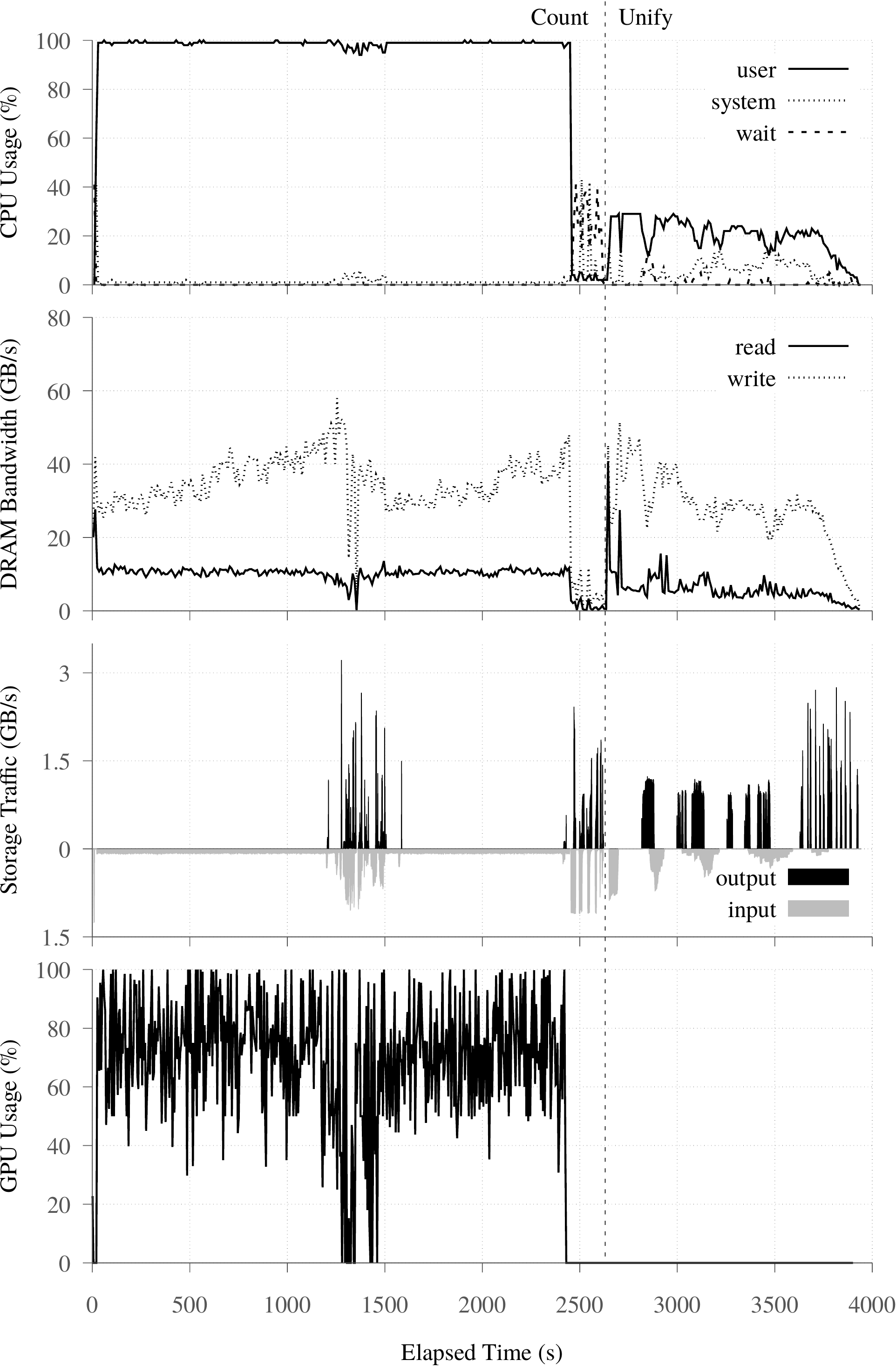}
\caption{Traces of the system when running Count and Unify units.
Prune unit is omitted for space constraints and similarity to
the Count unit.}
\label{f_traces}
\end{figure}

Traces in Figure~\ref{f_traces} show how the CPU is the main
bottleneck for most of the Count unit while the GPU is not fully utilized.
In this unit, in fact, we can see a steady 100\% utilization of the CPU aside for
a feeble reduction in the middle and last parts.
At these points in time, the application is swapping k-mer tables
to the NVM resulting in IO bursts in the storage traffic plot.
From the IO bursts happening in the middle of the Count unit we can see how the CPU,
getting busier to do the IO, further decreases the GPU usage; justifying the
simultaneous drop in the GPU utilization.
The second and last group of IO bursts in the Count unit represent the swapping of the tables
at the end of the input and by this time the GPU is already released which explains
the total absence of GPU activity.
Although for the first IO bursts there is a reduction in the CPU usage for the
second instead, the wait CPU time shows peaks at 40\%.
This difference is due to the fact that while our swapping mechanism can be set
to mitigate the IO bursts, at the end of the count there is nothing to process so
we do not mind to flood the NVM with 48 threads writing.
The lower CPU usage in the Unify unit is mainly due to the IO intensive nature
of this unit.
Because of it, in this unit, we only use a dozen of CPU threads which are already
able to put enough pressure on the IO subsystem.
Besides, empirical tests showed that a higher number of CPU threads is
counterproductive and rapidly exacerbates the CPU wait time.

\begin{figure}
\includegraphics[width=1.00\columnwidth]{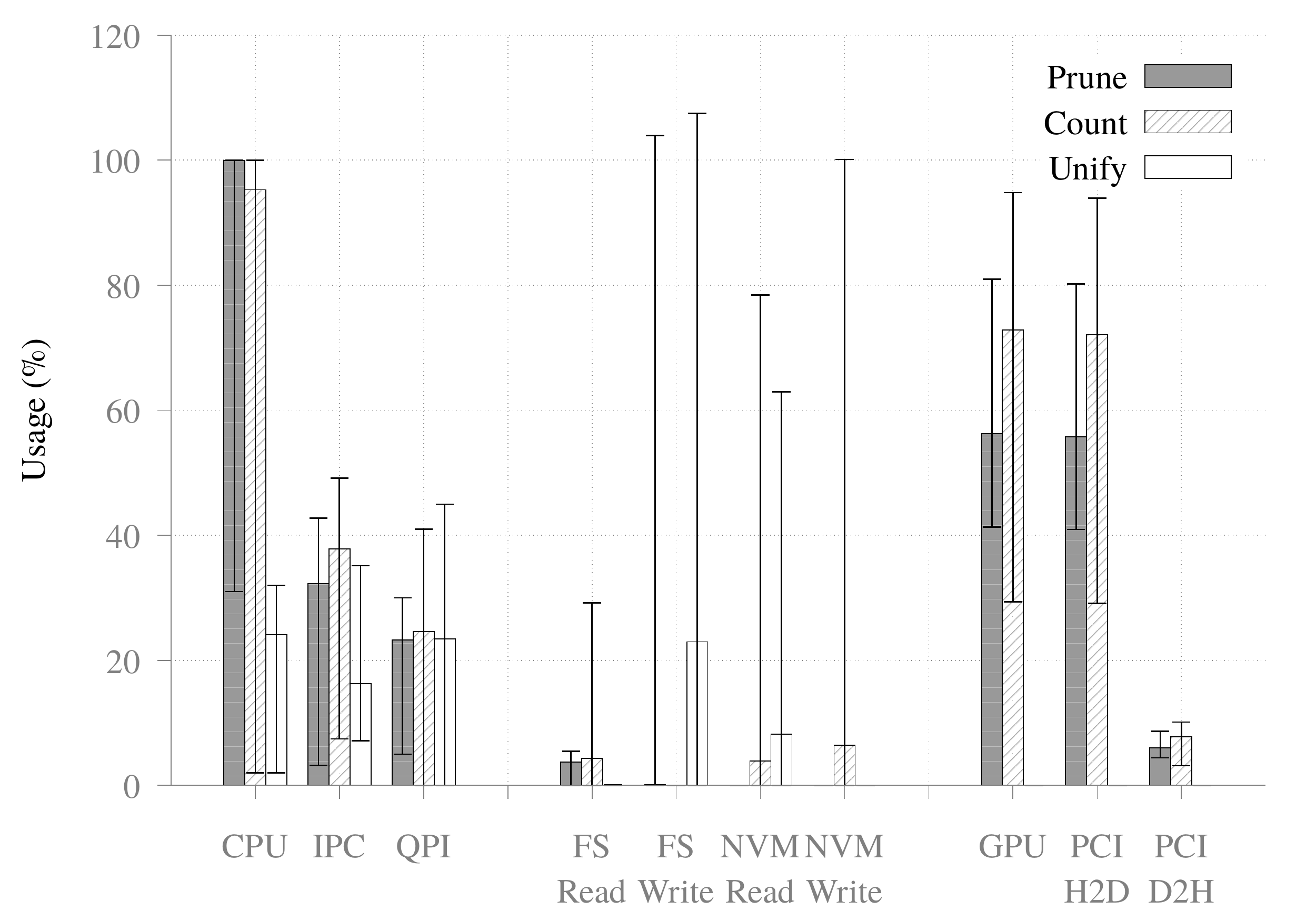}
\caption{Usage of major components in the system.
IPC is normalized to the maximum achievable values which for the used CPU
microarchitecture is 4.
QPI, FS and NVM drives are normalized to their maximum bandwidth.
Data relative to GPU and PCI represent the amount of time the resource was used over the total execution time
}
\label{f_resources}
\end{figure}

Figure~\ref{f_resources} shows additional information on the utilization of the
resources for all three units.
As expected it shows a high CPU usage for both Prune and Count units with low
minimum values characteristic of the end of each unit where the application is
check-pointing dumping the Bloom filters and the k-mer tables.
An activity that is the cause of the maximum values for the ``FS Write" and ``NVM
Write" in the Prune and Count unit respectively. Similarly the maximum value for
``FS Write" is the Unify unit is given by the dumping of stem tables.
Note that those values over 100\% depict to the ability of the NVM
drive to absorb short bursts of IO thanks to their on-board DRAM.

For Prune and Count units, Figure~\ref{f_resources} also unveils that,
even if fully utilized, the CPU is ineffective
and delivers a meager IPC (Instruction Per Cycle) throughout the
execution.
Further profiling showed that around one-third of CPU cycles are wasted in stalls
due to L1D cache misses and that up to half are instead spent waiting for the
memory subsystem.
Misses that are due to the small random updates in the Bloom filters and in the Tables.
To improve memory latency we tried all the three QPI (QuickPath Interconnect) snoop
protocol possible:
\textit{Home Snoop} (default), \textit{Early Snoop} and \textit{COD (Cluster-on-die)}.
As shown in~\cite{qpi} each protocol improves a different kind of workloads:
Home Snoop increases sustained bandwidth, Early Snoop shortens the latency
between sockets, and COD improves the latency for local memory accesses.
Results showed that whereas Early Snoop yielded a 5\%
improvement (6 minutes out of 114) in the execution time from Prune to Unify,
COD instead increased it of 48 minutes.

In conclusion, the memory latency should be considered the bottleneck and the
poor data locality of the memory accesses to the Bloom filters
and to the frequency tables is to be
blamed for the steady 100\% CPU utilization but poor IPC.
Moreover, considering the size of such data structures, larger CPUs caches
would provide hardly any benefits.

\section{Related Work}
\label{sec:related}

Counting k-mer frequencies is a widely studied problem in the literature and
different kinds of data structures have been used to solve it~\cite{laehnemann_denoising_2016}.
From enhanced suffix arrays in Tallymer~\cite{kurtz2008new},
lock-free hash tables in Jellyfish~\cite{btu071}, to more common hash tables,
Bloom filters, or a combination of the two like BFCounter~\cite{melsted2011efficient}.
However, most k-mer counters focus on either smaller values of $k$ or
smaller input sizes than required for \sm.
Moreover, some of such works also include specific optimizations to discard
the least frequent k-mers like the chain of Bloom filter implemented in \sm.

Offload part of genomic workloads with accelerators has also been studied before.
An example is BLAST, a local alignment search tool, which has been accelerated
adopting FPGAs as described in~\cite{datta_rc-blastn:_2009} and
in~\cite{krishnamurthy_biosequence_2004}.
In the latter case, authors also used offloaded Bloom filters to discard a large
fraction of a genome.
The same strategy ported to GPUs was instead implemented in~\cite{ma_bloom_2011}.
However, both cases used Bloom filters with smaller than 1 MB
while our implementation uses much bigger Bloom filters.
In~\cite{liu_decgpu:_2011} instead, counting Bloom filters are used
to obtain similar results, but this kind of Bloom filters has a
memory footprint that is 3 to 4 times higher than the normal filter.
And would exacerbate the issue of few GPU memory and cannot be afforded in our
application.

Similarly to our manual swapping mechanism to NVM other
k-mer counting methods like DSK~\cite{dsk} and Jellyfish~\cite{btu071}
also dump tables results to disk to keep a low DRAM memory
footprint.
However, they do not take advantage of the performance boost given by NVM.
Either they use few CPU Threads unable to saturate the drives or they used
rotational disks.

\section{Conclusions}
\label{sec:conclusions}

In this paper, we illustrated how the data intensive nature of counting k-mers in the
human genome is still a computational and memory challenge.
However, we described techniques and mechanisms to overcome the memory challenge
and to alleviate the computational one.
In particular, we have demonstrated how a GPU can be used to shuffle data
to minimize inter-thread communication and how it can cooperate with the CPU to build large
Bloom filters. While the former is important for vertical scalability, the
latter is used to unburden the CPU of some work.
Both of these novel methods were used in this work for k-mer counting but they
can be adopted to improve algorithms from other domains.
We have also illustrated how NVM can be used to swap hash tables from DRAM in a
controlled way to minimize CPU waiting time.
We applied this work to \sm, a real-world production genomics application that
performs a direct comparison of normal and tumor genomic samples from the same
patient. And we scaled-up the application to run in single node machine equipped
with a GPU and NVM drives that can be easily hosted in hospitals.

Results showed that although the single machine is not able to improve the time
to solution of 16 \mn nodes, the CPU time of the single machine is
7.5x shorter than the aggregate CPU time of the 16 nodes, with
a reduction in energy consumption of 5.5x.
In large supercomputing facilities, this work makes massive genome analysis
at an affordable energy consumption a possible reality.

We are currently working on a multi-GPU implementation, a way to improve data
locality, and also exploring how NVM sharing and disaggregation can reduce the
total cost of ownership.

\section*{Acknowledgments}
This project has received
funding from the European Research Council (ERC) under the
European Union's Horizon 2020 research and innovation programme
(grant agreement No 639595). It is also partially supported by the
Ministry of Economy of Spain under contract TIN2015-65316-P and Generalitat
de Catalunya under contract 2014SGR1051, by the ICREA Academia program,
and by the BSC-CNS Severo Ochoa program (SEV-2015-0493).

We are also grateful to SandDisk for lending the FusionIO cards and to Nvidia who donated the Tesla K40c.

%
%
%
\bibliographystyle{IEEEtran}
\bibliography{IEEEabrv,references.bib}

\end{document}